\documentclass[prl,twocolumn,superscriptaddress,amssymb]{revtex4}
\usepackage[dvips]{graphicx}
\usepackage{latexsym}
\usepackage{bm}

\begin{document}
\renewcommand{\ni}{{\noindent}}
\newcommand{\dprime}{{\prime\prime}}
\newcommand{\be}{\begin{equation}}
\newcommand{\ee}{\end{equation}}
\newcommand{\bea}{\begin{eqnarray}}
\newcommand{\eea}{\end{eqnarray}}
\newcommand{\la}{\langle}
\newcommand{\ra}{\rangle}
\newcommand{\dg}{\dagger}
\newcommand\lbs{\left[}
\newcommand\rbs{\right]}
\newcommand\lbr{\left(}
\newcommand\rbr{\right)}
\newcommand\f{\frac}
\newcommand\e{\epsilon}
\newcommand\ua{\uparrow}
\newcommand\da{\downarrow}
\title{Can correlations drive a band insulator metallic?}
\author{Arti Garg}
\affiliation{Department of Theoretical Physics, Tata Institute of
Fundamental Research, Mumbai 400005, India}
\author{H. R. Krishnamurthy}
\affiliation{Centre for Condensed Matter Theory, Department of Physics,
Indian Institute of Science, Bangalore 560 012,
and Condensed matter Theory Unit, JNCASR, Jakkur, Bangalore 560 064, India}
\author{Mohit Randeria}
\affiliation{
Department of Physics, The Ohio State University, Columbus, OH 43210}
\vspace{0.2cm}
\begin{abstract}
\vspace{0.3cm}
We analyze the effects of the on-site Coulomb repulsion $U$ on a band insulator
using dynamical mean field theory (DMFT). We find the surprising result that the gap
is suppressed to zero at a critical $U_{c1}$ and remains zero within a metallic phase.
At a larger $U_{c2}$ there is a second transition from the metal to a Mott insulator,
in which the gap increases with increasing $U$. These results are qualitatively different from
Hartree-Fock theory which gives a monotonically decreasing but non-zero insulating gap
for all finite $U$.
\vspace{0.1cm}
\typeout{polish abstract}
\end{abstract}
\maketitle

There are many examples of interactions opening a gap in a
metallic state: the Mott metal-insulator transition, as well as
spin or charge density wave gaps on nested Fermi surfaces.
However, the possibility
of interactions suppressing the gap in a band
insulator all the way down to zero seems, at first sight, to be
counterintuitive. In this Letter, we show that this does occur in a
simple model treated within a well-tested
approximation scheme, namely dynamical mean field theory
(DMFT)~\cite{georges,jarrell}.

\begin{figure}
\begin{center}
\includegraphics[width=2.75in,angle=0]{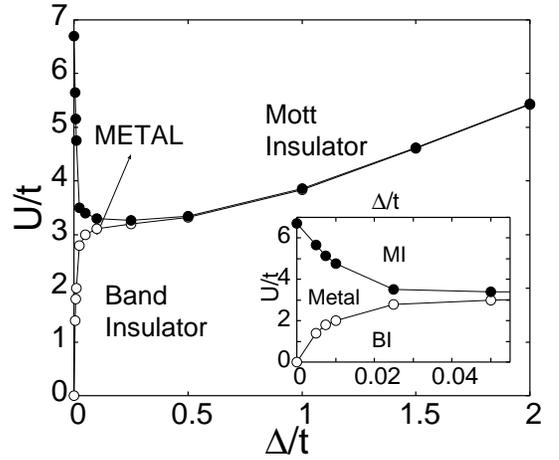}
\caption{The $T=0$ phase diagram of model of eq.~(1) at
half-filling obtained using DMFT for the Bethe lattice. The 
phase boundary $U_{c1}$ is denoted by open circles while
$U_{c2}$ by filled circles. Inset: Detail
of lower left hand corner of the phase diagram.}
\label{phase_diag}
\end{center}
\vskip-6mm
\end{figure}

Specifically, we start with a simple tight-binding band insulator
with two bands, one filled and the other unfilled, and turn on an
on-site Coulomb repulsion, the Hubbard $U$. We treat the effect of
$U$ using the DMFT within the paramagnetic sector, and show that
with increasing $U$ the gap of the band insulator first collapses
to zero at a $U_{c1}$ and remains zero,
leading to a metallic phase, for a finite range of $U$. This may
be intuitively understood in terms of correlations screening out
the one-body potential responsible for the gap in the band
insulator. Eventually, with a further increase in $U$,  there is a
second phase transition at a $U_{c2}$ to a Mott insulator. Our
calculated phase diagram is shown in Fig.~\ref{phase_diag}.

The model we consider has tight-binding electrons on a bipartite
lattice (sublattices A and B) described by
\[
H=-t\sum_{i\in A,j\in B,\sigma} [~c^{\dagger}_{i\sigma}c_{j\sigma}
+h.c~]+ \Delta \sum_{i\in A}n_{i} -\Delta\sum_{i \in B} n_{i} \]
\be
\mbox{~~~~~~~~~~~~~~~~~~~~~~~~~~~}+U\sum_{i}n_{i\ua}n_{i\da}-\mu
\sum_{i}n_{i} \label {model} 
\ee 
where $t$ is the nearest neighbor
hopping, $U$ the Hubbard repulsion and $\Delta$ a one-body
potential which doubles the unit cell. The chemical potential is
chosen \cite{phsymm} to be $\mu=U/2$, so that the 
average occupancy is $\left(\langle n_A
\rangle + \langle n_B \rangle \right)/2=1$
(``half-filling''). The Hamiltonian (\ref
{model}) is sometimes called the ``ionic Hubbard model'' with
$\Delta$ the ``ionic'' potential, and has been extensively studied
especially in one
dimension~\cite{hubbard,egami,ortiz,resta,fabrizio,wilkens,aligia},
although the existence of an intermediate metallic state between
the band and Mott insulators is apparently not settled
conclusively even in the 1D case. In any case, both our motivation
and methodology are very different from previous
studies.

We begin by considering various limiting cases at half filling. In
the non-interacting ($U=0$) limit the system is a band insulator
with a band gap of $\Delta$. However, a very different picture
emerges in the atomic limit ($t=0$). Here we obtain a ``band
insulator'' for $U/2 < \Delta$ with site occupancies $n_A = 0$ and
$n_B = 2$ and a gap $\Delta-U/2$ for either injecting or
extracting an electron. This gap decreases with increasing $U$ and
collapses to zero at $U/2 = \Delta$ when the singly-occupied
states becomes degenerate with the empty one at site $A$ and with
the doubly-occupied one at site $B$. Beyond this point, for $U/2 >
\Delta$ we get a ``Mott insulator'' with $n_A = n_B = 1$, and a
gap that goes like $U$ when $U \gg \Delta$. Hence, in this
trivially soluble $t=0$ limit, we clearly see that the interaction
$U$ suppresses the gap of (the state adiabatically connected to)
the band insulator all the way down to zero, but only at a single
point $U = 2\Delta$.
A question of considerable interest is whether this metallic {\it
point} obtained in the atomic limit
can broaden into a metallic \emph{phase} when $t$ is non-zero.

Here we address this question using the DMFT approach.
The DMFT approximation is exact in the limit of large
dimensionality \cite{georges,jarrell} and has been demonstrated to be very
successful in understanding the metal-insulator transition
\cite{georges,jarrell} in the usual Hubbard model, which is the
$\Delta =0$ limit of eq.~(\ref{model}). We focus in this paper on
the paramagnetic sector of
eq.~(\ref{model}), for which it is convenient to introduce the
matrix Green's function \be
\hat{G}_{\alpha\beta}({\bf{k}},i\omega_n)
= \lbr \begin{array} {cc} \zeta_{A}({\bf{k}},i\omega_n) & -\epsilon_{\bf{k}} \\
-\epsilon_{\bf{k}} & \zeta_{B}({\bf{k}},i\omega_n)
\end{array}\rbr^{-1} \label{Greensfn} \ee where $\alpha,\beta$ are
sublattice ($A,B$) indices, ${\bf k}$ belongs to the first
Brillouin Zone (BZ) of \emph{one sublattice},
$i\omega_n=(2n+1)\pi T$ and $T$ is the temperature.
The kinetic energy is described by the dispersion
$\epsilon_k$ and $\zeta_{A(B)} \equiv i\omega_n \mp
\Delta+\mu-\Sigma_{A(B)}(i\omega_n)$. Within the DMFT approach the
self energy is purely local~\cite{georges}. Thus the diagonal
self-energies $\Sigma_\alpha(i\omega_n)$ are ${\bf
k}$-independent and the off-diagonal self-energies vanish 
(since the latter would couple the A and B sublattices).

We note that within the \emph{Hartree-Fock} (HF) approximation the
self energy is $\Sigma_{\alpha}^{HF}=U \langle n_\alpha \rangle/2$
and the the difference in filling factors of the two sublattices
$\delta n = \left(\langle n_B \rangle - \langle n_A \rangle
\right)/2$ is given by the $T=0$ self-consistent equation $\delta
n = (\Delta-U \delta n/2) \sum_{\bf{k}} E_{\bf{k}}^{-1}$, where
$E_{\bf{k}} = \sqrt{\epsilon_{\bf{k}}^2+(\Delta-U\delta n/2)^2}$
is the HF excitation spectrum. Thus, within HF theory there is a
gap $(\Delta-U\delta n/2)$ for all $\Delta \ne 0$ and the system
is a band insulator for all $U$. (For $\Delta=0$, one gets $\delta
n=0$ and the system is a metal for all $U$). However,
a much richer, qualitatively different phase diagram emerges
when we include fluctuations beyond HF, even at the level of the
DMFT.

The DMFT approach includes {\it local} quantum fluctuations by
mapping \cite{georges,jarrell} the lattice problem onto a
single-site or ``impurity'' with local interaction $U$
hybridizing with a self-consistently determined bath as follows.
(i) We start with a guess for $\Sigma_\alpha(\omega^{+})$ and
$\delta n$ and compute the local
$G_\alpha(i\omega_{n})=\sum_{{\bf{k}}}G_{\alpha\alpha}({\bf{k}},i\omega_{n})$
rewritten as \be
G_{\alpha}(i\omega_n)=\zeta_{\bar{\alpha}}(i\omega_n)\int_{-\infty}^{\infty}
d\epsilon
\frac{\rho_{0}(\epsilon)}{\zeta_{A}(i\omega_n)\zeta_{B}(i\omega_n)-\epsilon^{2}}
\label{fullG} \ee with $\alpha=A(B)$ and $\bar{\alpha}=B(A)$ with
$\rho_{0}(\epsilon)$ is the bare density of states (DOS) for the
lattice considered (see below). We actually need to solve the
problem for only one sublattice and use the relations
$G_{A}(i\omega_n) = -G_{B}(-i\omega_n)$ and $\Sigma_{A}(i\omega_n)
= U-\Sigma_B(-i\omega_n)$ valid at half filling. (ii) We next
determine the ``host Green's function"~\cite{georges,jarrell}
$\mathcal{G}_{0\alpha}$ from the Dyson equation
$\mathcal{G}_{0\alpha}^{-1}(i\omega_n) =
G_{\alpha}^{-1}(i\omega_n) + \Sigma_{\alpha}(i\omega_n)$. (iii)
We solve the impurity problem to obtain
$\Sigma_\alpha(i\omega_{n}) = \Sigma_\alpha
\left[\mathcal{G}_{0\alpha}(i\omega_{n})\right]$
(iv) We iterate steps (i), (ii) and (iii) till a self-consistent
solution is obtained.

We use as our ``impurity solver'' in step (iii) a generalization
of the iterated perturbation theory (IPT) \cite{georges,kk} scheme
which has the merit of giving semi-analytical results directly in
the real frequency ($\omega^+ = \omega + i0^+$) domain. The IPT
ansatz
$\Sigma_{\alpha}^{IPT}(\omega^{+})=\Sigma_{\alpha}^{HF}+A_{\alpha}\Sigma_{\alpha}^{(2)}(\omega^{+})$
is constructed to be (a) exact for $U/t \ll 1$, (b) exact for
$t/U=0$, and (c) exact in the large $\omega$ limit for all $U/t$,
which imposes various exact sum rules. Here $\Sigma_\alpha^{HF}$
is the HF self energy with $n_{\alpha}=-2 \int_{-\infty}^{0}
\mbox{Im~} G_{\alpha}(\omega^{+}) d \omega/\pi$, and 
\be
\Sigma_{\alpha}^{(2)}(\omega^{+})=U^{2}\int_{-\infty}^{\infty}
\prod_{i=1}^{3} \left[ d \epsilon_{i} \tilde{\rho}_{\alpha}(\epsilon_{i}) \right]
\f{N(\epsilon_{1},\epsilon_{2},\epsilon_{3})}
{\omega^{+}-\epsilon_{1}+\epsilon_{2}-\epsilon_{3}}
. \label{sigma2} 
\ee 
This has the form of the second order self-energy with
$\tilde{\rho}_{\alpha}(\epsilon_{i})=-\mbox{Im}[\tilde{\mathcal{G}}_{0
\alpha}(\epsilon_{i}^{+})]/\pi$, where
$\tilde{\mathcal{G}}_{0\alpha}^{-1}(\omega^{+})=
\mathcal{G}_{0\alpha}^{-1}(\omega^{+})-\Sigma_{\alpha}^{HF}$ is
the Hartree corrected host Green's function and
$N(\epsilon_{1},\epsilon_{2},\epsilon_{3})=f(\epsilon_{1})f(-\epsilon_{2})f(\epsilon_{3})
+f(-\epsilon_{1})f(\epsilon_{2})f(-\epsilon_{3})$ where
$f(\epsilon)$ is the Fermi function. From condition (c) above we
find that $A_{\alpha}=
n_{\alpha}(1-n_{\alpha}/2)/\left[n_{0\alpha}(1-n_{0\alpha}/2)\right]$
with $n_{0\alpha}=-2\int
_{-\infty}^{0}\mbox{Im}\tilde{\mathcal{G}}_{0\alpha}(\omega^{+})d
\omega/\pi$. Note that $A_\alpha$ is same for both the
sublattices. The results of this simple approximate method are
expected to be in semi-quantitative agreement with those of more
exact but numerically intensive methods 
\cite{georges,jarrell,bulla}. For
simplicity, here we present the results for the $T=0$ solution of
DMFT equations on a Bethe lattice of connectivity $z \rightarrow
\infty$.
The hopping amplitude is rescaled as $t\rightarrow t/\sqrt{z}$ to
get a non-trivial limit and the bare DOS is then given by
$\rho_0(\epsilon) = \sqrt{4t^2-\epsilon^2}/(2\pi t^2)$ which
greatly simplifies the integral in eq.~(\ref{fullG}).
We have also solved the DMFT equations on the 2D square lattice
and found that the results are qualitatively the same\cite{garg}.

\begin{figure}[h!]
\begin{center}
\includegraphics[width=2.25in,angle=0]{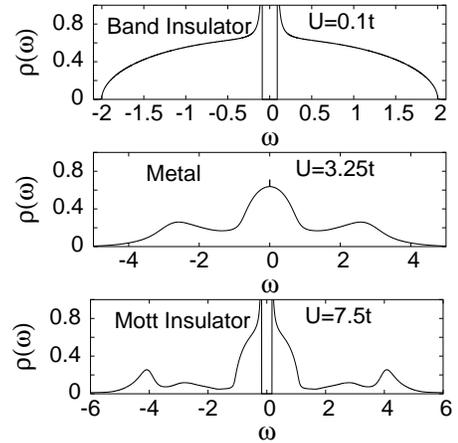}
\caption{The density of states $\rho(\omega)$ (in units of $1/t$)
plotted as a function of $\omega$ (in units of $t$). 
The three panels are all at $\Delta=0.1t$
for various values of $U$ one for each phase. The gaps are clearly
visible in the top (band insulator at $U=0.1t$) and bottom (Mott
insulator at $U=7.5t$) panels, while the middle panel for metallic
state at $U=3.25t$ shows gapless excitations.} \label{dos_bethe}
\end{center}
\vskip-6mm
\end{figure}

The phase diagram of Fig.~\ref{phase_diag} has been obtained from
the Green's function, and the (renormalized) density of states
(DOS), given by $\rho(\omega) = -\sum_{\bf k}{\rm Im
Tr}[\hat{G}({\bf k},\omega^{+})]/\pi$, calculated using the above
procedure. Although the sublattice DOS $\rho_\alpha$ are not
individually symmetric, $\rho_B(\omega)=\rho_A(-\omega)$ and thus
$\rho(\omega)=\rho_A(\omega)+\rho_B(\omega)$ is symmetric for half-filling. 
Fig.~\ref{dos_bethe} shows
how $\rho(\omega)$ evolves as a function of $U$ for a fixed
$\Delta$. At small $U < U_{c1}$
there is a gap in the spectrum, with the DOS at higher energies
similar to the non-interacting (semicircular) result. We call this
state a band insulator since it is adiabatically connected to the
$U=0$ band insulator. At intermediate $U$ the gap collapses
to zero \cite{numerical_checks}, and we find a metallic state.
The effects of correlations also show up in the precursors to the
upper and lower Hubbard bands at higher energies. At sufficiently
large $U > U_{c2}$, the DOS again shows a gap and the system is a
Mott insulator (adiabatically connected to the well studied Mott
insulator at $\Delta = 0$).
The metallic phase, sandwiched between $U_{c1}(\Delta)$ and
$U_{c2}(\Delta)$ in Fig.~\ref{phase_diag}, shrinks as $\Delta$
increases, consistent with
there being a single metallic `point' $U = 2\Delta$ in the atomic
limit. We return below to the question of the scale
of the critical $U_{c}$'s as a function of $\Delta$.

\begin{figure}[h!]
\begin{center}
\includegraphics[width=2.0in,angle=-90]{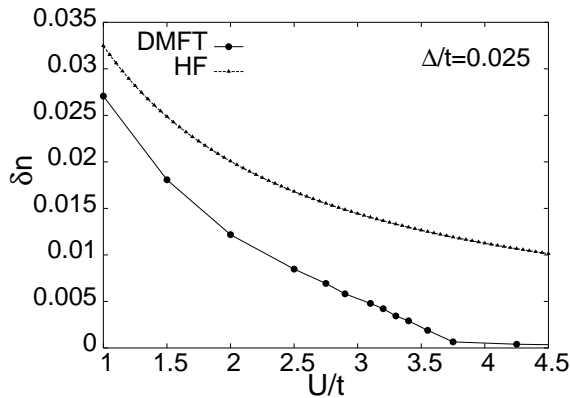}
\caption{Difference in filling factor $\delta n = (n_B-n_A)/2$ for
$\Delta=0.025t$ plotted as a function of $U/t$:
The DMFT result (filled circles) is much smaller
than the HF result (triangles).
} 
\label{deltan}
\end{center}
\vskip-3mm
\end{figure}

\begin{figure}[h!]
\begin{center}
\includegraphics[width=2.0in,angle=-90]{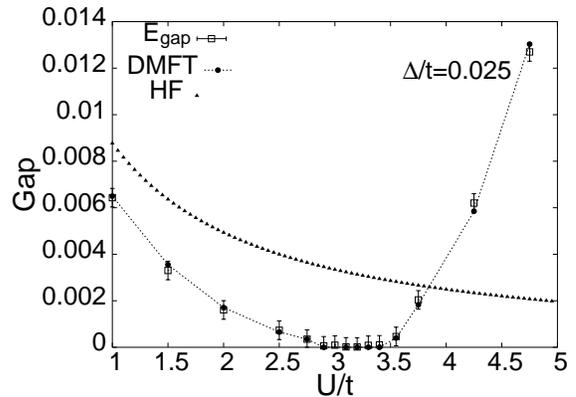}
\caption{The energy gap for $\Delta=0.025t$ plotted as a function
of $U/t$. The DMFT gap (dots) from the DOS and $E_{\rm gap}$ 
(open squares) from eq.~(\ref{gap}) are in excellent agreement. Note that 
the gap vanishes in the metallic phase at intermediate $U$'s. 
(See text for discussion of error bars). In contrast, the HF results (triangles) 
do not show a metallic phase.}
\label{Gap_HF_DMFT}
\end{center}
\vskip-3mm
\end{figure}

The difference in filling $\delta n=(n_B-n_A)/2$ and the 
energy gap are plotted in Fig.\ref{deltan} and Fig.~\ref{Gap_HF_DMFT}
as functions of $U$ for fixed $\Delta$. For $\delta n$ the
DMFT results are much smaller than the HF results
and become very small (of the order of error in evaluating $\delta
n$)for large values of U. But in the large U limit $U\gg
t,\Delta$, one can map the model (1) to the Heisenberg
model with the spin exchange integral $J=4t^2U/(U^2-4\Delta^2)$
and show that
$\delta n\approx 16t^2\Delta/U^3$ is non zero.
The DMFT result for the energy gap is qualitatively different from
the simple HF result (monotonically decreasing, non-zero for all
$U$), and vanishes continuously\cite{numerical_checks} at both
metal-insulator transitions, at $U_{c1}$ and $U_{c2}$.

\begin{figure}[h!]
\begin{center}
\includegraphics[width=2.25in,angle=0]{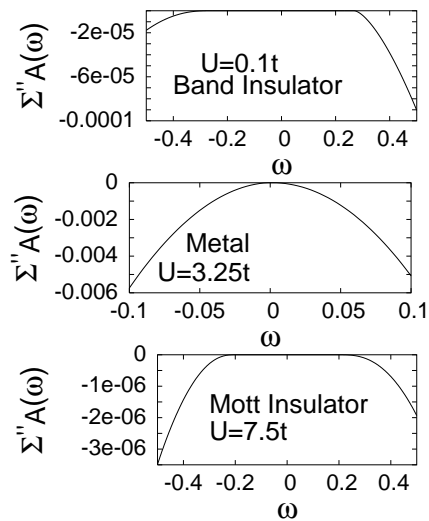}
\caption{Imaginary part of self energy
$\Sigma_A^{\prime\prime}(\omega)$ plotted as a function of
$\omega$, where both are measured in units of $t$. All three
panels are for $\Delta=0.1t$ for various values of $U$. In the top
(band insulator at $U=0.1t$) and bottom (Mott insulator at
$U=7.5t$) panels $\Sigma_A^{\prime\prime}(\omega)$ vanishes
identically for low energies $\vert \omega \vert < 3E_{\rm gap}$.
The metallic phase in the middle panel at $U=3.25t$ has a
Fermi-liquid-like $\Sigma_A^{\prime\prime}(\omega) \sim \omega^2$. }
\label{sigma_bethe}
\end{center}
\vskip-3mm
\end{figure}

Further insight into the scale of the gap and its non-monotonic
variation with $U$ is obtained by examining the self energy
$\Sigma_\alpha(\omega^{+})= \Sigma_\alpha^{\prime}(\omega)+i
\Sigma_\alpha^{\prime\prime}(\omega)$. From Fig.~\ref{sigma_bethe}
we see that $\Sigma^{\prime\prime}_A(\omega)$ vanishes for
$|\omega|\le 3E_{\rm gap}$ in both the insulating phases. This can
be understood from of the imaginary part of $\Sigma_\alpha^{(2)}$
of eq.~(\ref{sigma2}), which comes from a three fermion
final state. In the metallic phase
$\Sigma_A^{\prime\prime}(\omega)\sim \omega^2$ near $\omega=0$,
characteristic of a Landau Fermi liquid. In all the three phases,
$\Sigma_A^{\prime}(\omega)$ 
can be written at low $\omega$ as a Taylor expansion
$\Sigma_\alpha^{\prime}(\omega)=\Sigma_\alpha^{\prime}(0)+ \lbr 1-
Z^{-1}\rbr \omega + \ldots$, where $Z$ can be shown to be
independent of $\alpha$. The spectral function is
defined by $\mathcal{A}_{\alpha\alpha}(\epsilon,\omega)=-1/\pi
{\rm Im} G_{\alpha\alpha}(\epsilon,\omega^{+})$. When
$\Sigma_\alpha^{\prime\prime}$ is negligible we find, from
eq.~(\ref{Greensfn}), that $A_{\alpha\alpha}(\epsilon,\omega) =
\delta(r(\omega)-\epsilon^2)$ with
$r(\omega)=(\omega+\mu-\Delta-\Sigma^{\prime}_{A}(\omega))(\omega+\mu+\Delta-\Sigma^{\prime}_{B}(\omega))$.
As $\epsilon$ is real, $\omega$'s which satisfy $r(\omega) < 0$
lie within the gap. The energy gap is then given by $r\left(E_{\rm
gap}\right)= 0$ which, using the low-energy form of
$\Sigma_\alpha^{\prime}$ given above and the Kramers-Kronig
relation, leads to the result \be E_{\rm gap} = Z\vert
\Delta-{U\delta n / 2}+ S \vert \label{gap} \ee where $S =
P\int_{-\infty}^{\infty}d\omega
\Sigma_A^{\prime\prime}(\omega)/\pi\omega$. We note that the
energy gap obtained from the DOS (filled circles in
Fig.~\ref{Gap_HF_DMFT}) is in excellent agreement with that given
by eq.~(\ref{gap}) (open squares in Fig.~\ref{Gap_HF_DMFT}), where
the error bars on the latter are obtained from the estimated
numerical error in $\Sigma^{\prime\prime}$ at small $\omega$
\cite{numerical_checks}. In particular we have checked very
carefully that the gap indeed vanishes in the entire metallic
phase within the limits of our numerical accuracy.

We note that Eq.~(\ref{gap}) helps greatly in clarifying several
features of our results\cite{garg}. First, it  shows explicitly
how correlations ``screen'' the one-body potential $\Delta$ responsible
for the gap in the band insulator and lead to a gap which is
always less than the HF result in the band insulating phase, The
dominant role in this being played by $S (< 0)$ and by $Z (< 1)$.
Second, we estimate the location $U_{c1}$ of the band insulator to
metal transition using $E_{\rm gap} = 0$ which implies $U = 2\vert
\Delta + S(U) \vert / \delta n (U)$. We find from our numerics
that although $\vert S(U)\vert$ increases with $U$, it is always
much smaller than $\Delta$. Thus $U_{c1} \simeq 2 \Delta / \delta
n(U_{c1}) \ge 2 \Delta / \delta n(0)$, since $\delta n(U)$ decreases 
with increasing $U$ (see Fig.~3). Thus for
small $\Delta$, we find that the very small $\delta n$ leads to
$U_{c1} \gg \Delta$; e.g., for $\Delta=0.01t$ we get $U_{c1} \ge
100\Delta \simeq t$, while for $\Delta=0.025t$, $U_{c1} \ge
20\Delta \simeq 0.5 t$. Finally, the eventual increase in the gap
with $U$ in the Mott phase, can be also traced to the
$U$-dependences of $Z$ and $S$ in eq.~(\ref{gap}).

We find that \cite{garg} the low energy spectral function in the metallic
phase $\mathcal{A}(\epsilon,\omega)=-{\rm Im
Tr}[\hat{G}(\epsilon,\omega^{+})] \simeq
Z\delta(\omega-Z\epsilon)$. Thus the Fermi Surface is given by
$\epsilon = 0$, i.e., the same as for the metal with $U = \Delta
=0$, as a consequence of the ${\bf{k}}$-independence of DMFT self
energy. The quasiparticle residue $Z$, which is also the
mass-renormalization, is a function of both $U$ and
$\Delta$. Within the metallic phase it decreases with increasing
$U$ for a given $\Delta$, and with increasing $\Delta$ for fixed
$U$.

In conclusion, we have analyzed a simple model of a band insulator
with on-site Coulomb repulsion $U$. At the Hartree-Fock level the
gap is reduced but the insulating behavior persists. However, when
we treat correlation effects using DMFT we find the surprising
result that the gap is suppressed to zero and there is a band
insulator to metal transition at a critical value of $U_{c1}$. At
an even larger $U_{c2}$ there is a second transition from the
metal to a Mott insulator.

Our work raises several important questions. First, our results
are obtained in the paramagnetic sector. We expect that the
metallic phase will survive if antiferromagnetism~\cite{garg} 
is suppressed due to frustration, in much the
same way as in the DMFT treatment in the $\Delta = 0$
limit\cite{georges}.
It would be interesting to treat a model in which such frustration is explicitly  included.
Second, it is important to ask if in finite dimensions, i.e., with
${\bf k}$-dependent self energies, other broken symmetry states
might also appear, for example the bond ordered state which has
been proposed as an intermediate phase in low dimensional
studies~\cite{fabrizio,wilkens,aligia}. Finally, it would be most
interesting to look for experimental systems -- either transition
metal oxides \cite{imada_rmp} or even fermions in optical lattices
\cite{esslinger} where increasing correlations could drive a
band-insulator metallic. All these are questions for future work.

MR would like to thank N. Trivedi for very useful conversations.


\begin{thebibliography}{99}
\bibitem{phsymm}
This can be shown by using particle-hole symmetry at half-filling.
\bibitem{numerical_checks}
The very low energy results for the DOS and
$\Sigma^{\prime\prime}$ are highly susceptible to numerical
errors, particularly in the metallic phase.
We have checked that our results are robust under
increased numerical accuracy. In many cases we have also compared
results using two different technique (fast Fourier transform
and direct numerical integration). This forms the basis for
error bars on $E_{\rm gap}$ in Fig.~3. For reasons related to
numerical accuracy, we have not yet analysed the
possibilities of first order transitions near $U_{c2}$ which may
be present for small $\Delta$.
\end{thebibliography}
\end{document}